\documentclass[fleqn,12pt]{wlscirep}
\usepackage[utf8]{inputenc}
\usepackage{amsmath}
\usepackage{physics}
\usepackage[T1]{fontenc}
\title{Tailoring the Down-Conversion Emission Profile Via Direct Imaging with Camera}
\author[1]{Hashir Kuniyil}
\author[1, *]{Kadir Durak}
\affil[1]{Dept. of Electrical and Electronics Engineering, Ozyegin University, Istanbul 34794, Turkey}
\affil[*]{kadirac@gmail.com}
\begin{document}

\begin{abstract}
We present the analysis of down-conversion emission profile from a critically phase matched type-I nonlinear crystal. This is done via direct imaging of down-converted photons by using a CMOS camera. Cylindrical asymmetry is observed in the emission profile due to the birefringent effects of the nonlinear crystal, i.e. pump walk-off. We provide a mathematical expression to theoretically recreate the down-converted mode shape. The experimentally observed mode and theoretically recreated one perfectly matches. We provide a method to eliminate the cylindrical asymmetry and correct it in the emission profile. Our method improves the collection of the down-converted photons to a single mode fiber by a factor of 2 compared to uncorrected case.  With the introduced method, we achieved 0.517 million coincidence events (per second per mW pump power).
\end{abstract}
\flushbottom
\maketitle
\section*{Introduction}
Photons are the most widely used particles for entanglement experiments due to their well-established production procedures and ease of manipulation in various types of media. Spontaneous parametric down-conversion (SPDC) is the most well-known technique for creating entangled photon pairs \cite{burnham1970observation,hong1986experimental,paulk}, in which the field of a pump photon is absorbed and two photons of lesser energy are re-emitted called signal and idler photons leaving the atoms of the crystal unaffected. The newborn daughter photons possess nonlocal correlations due to the energy and momentum conservation during the SPDC process. The pairs are created using either critically phase matched bulk crystals or more recently quasi-phase matched (or non-critically phase matched) crystals \cite{fejer1992quasi,fiorentino2007spontaneous,kwiat1999ultrabright}.

The photon pair sources are highly exploited in quantum key distribution \cite{kuzucu2005two}, quantum teleportation \cite{ma2007quantum} and quantum computation\cite{raussendorf2001one} experiments which utilizes quantum entanglement. This allowed highly successful demonstrations in the laboratory environments\cite{bouwmeester1997experimental}. However, it has only recently been studied to develop sources that can operate in rough conditions like space, underwater and various types of moving vehicles \cite{bennett1993teleporting,liao2017satellite,ji2017towards}. Such studies point out the strict considerations on the size, weight and power (SWaP) by keeping the brightness, collection efficiency and the entanglement fidelity maximum, where the brightness can be defined as the photon pair rate per second per mW pump power. Recently, robust and compact sources are developed that can even survive through rocket explosion \cite{tang2016photon,durak2016next}. However, it is still an ongoing quest to optimize the pair sources for having maximum brightness with desired entanglement quality to achieve wide range quantum key distribution networks\cite{zhang2018large}. In order to use the photon pairs at distant locations for quantum communications, they are required to be transported either in free-space or in fibers. For both cases the pairs are needed to be single (fundamental) Gaussian mode for their efficient use. Collecting the photons to a single mode fiber (SMF) can guarantee the single mode profile, and therefore easy manipulation of the photons. However, this spatial filtering reduces the brightness, which is very critical for the design process of the pair sources. For long crystals the pump walk-off causes asymmetries in the down-conversion emission profile due to the birefringence of the material. This increases the losses for the collection of the down-converted photons into a SMF.

In SPDC photon pair generation, each signal photon generated is correlated with each idler photon. Because of the experimental limitations, such as transverse momentum distribution of the pump and length of the crystal, the generated photon pairs have a definite spectral width. As a result, a Gaussian SPDC profile is observed for collinear geometry and an annular ring shaped mode for noncollinear geometry. This effect coupled with spacial walk-off effect in the birefringent nonlinear medium make it difficult to demonstrate an efficient SPDC sources. To reduce this problem, the short crystal approximation has been used in the context of many entangled photons based applications \cite{monken1998transfer,molina2005control,walborn2007transverse}. In this paper, we directly observe the down-converted photons using a complementary metal oxide semiconductor (CMOS) camera. This is a cheaper method compared to traditional imaging technique where a charge coupled device (CCD) is used to monitor the SPDC mode shape. The effects of SPDC parameters such as nonlinear crystal length, pump beam waist and the phase matching angle on collinear and noncollinear geometries have been studied separately. Our  camera observations suggest that the photon pair rate increases linearly with the crystal length as expected\cite{septriani2016thick}. Earlier investigations on the absolute down-conversion rates with finite pump beam waist values have shown that the rate of photon pair production is independent of the pump size\cite{kleinman1968theory}. On the other hand, the experimental results indicated some optimal pump beam size\cite{ljunggren2005optimal,pires2011type,ramirez2013effects}. Our findings show that the contradictions between these studies arise from the fact that the brightness is not independent from the collection optics. Our observed SPDC mode also shows a cylindrical asymmetry in its annular profile. We provide a simplified mathematical model to address this effect. The direct observation of the down-converted photons using a camera allows the experimental verification of the developed theoretical model. By monitoring the SPDC modes on camera, the collection mode is prepared to maximize the brightness. Brightness values of above 200 KCoincidences/mW/s can be achieved. We also introduce a method to eliminate the cylindrical asymmetry in the SPDC profile. With this method, the brightness of the SPDC after coupling to SMF can be easily doubled.
\section*{Theory}
In SPDC process, higher frequency pump photon produces two lower frequency photons called signal and idler photons. Energy and momentum conservation are the basic requirements to be satisfied to generate the target photons. In SPDC experiment, properties of nonlinear crystal is exploited to achieve target modes. More specifically, SPDC utilizes second order nonlinear susceptibility tensor $\chi^{\left(2\right)}$ in a suitable material. In our study, we use type-I negative $\beta$-barium borate (BBO) crystal as the nonlinear medium. In type-I negative crystals, the extraordinary polarized pump beam creates two ordinarily polarised SPDC photons ($ e\rightarrow o+o$). In the coming sections, we will use index $p$ for pump, index $s$ for signal and index $i$ for idler. In a customized nonlinear crystal, the desired wave vectors of signal and idler photons are generated. The SPDC target modes form a cone with its central axis in a direction with reference to the pump wave vector. Which means SPDC emission can occur at any angle defined by the input pump propagation direction. The phase matching of interaction fields in three dimension can be decomposed into,    
\begin{align}
  \Delta{k_z} = k_{p}-k_{s} \cos \theta_{s} +k_{i} \cos \theta_{i}
\end{align}
\begin{align}
  \Delta{k_x} = k_s \sin \theta_s - k_i \sin \theta_i
\end{align}
\begin{align}
  \Delta{k_y} = -k_s \sin \theta_s - k_i \sin \theta_i
\end{align}
where $k_{p}$, $k_{s}$ and $k_{i}$ are wave vectors of pump, signal and idler, respectively. $\theta_{s}$ and $\theta_{i}$ are the emission angles of signal and idler photons with respect to pump beam. For collinear geometry, the emission angles ($\theta_{s}, \theta_{i} $) become zero whereas, for noncollinear geometry $\theta_{s}, \theta_{i} $ are non-zero.\begin{figure}[h!]
\includegraphics[width=1\textwidth]{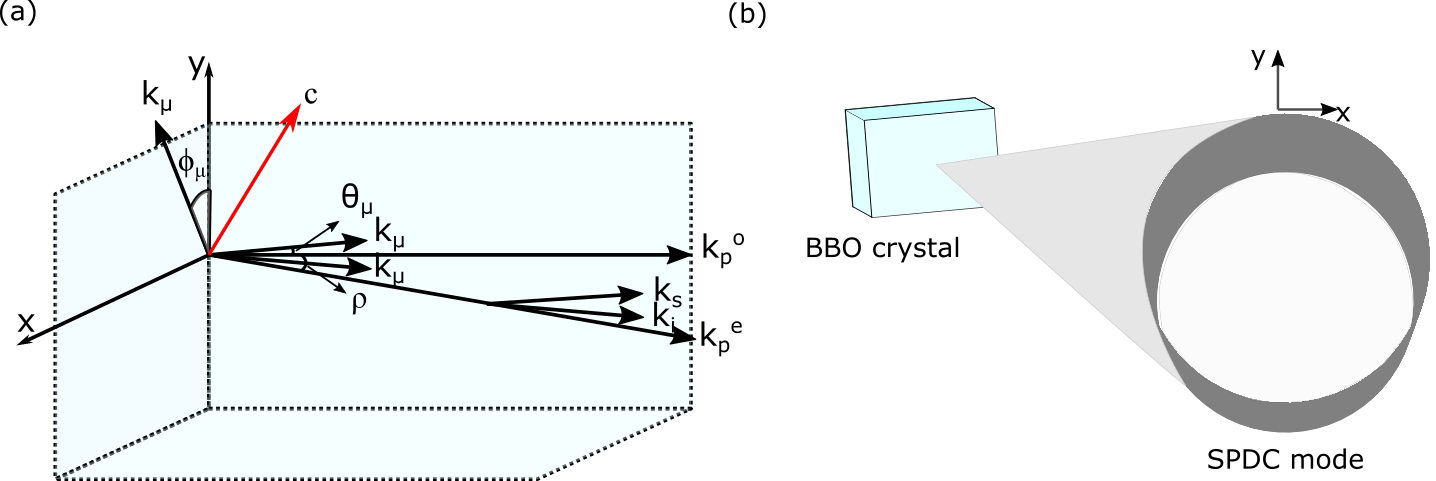}
\centering
\caption{Schematic of propagation direction of the pump, signal and idler beam with various source characteristics indicated, where $k_p^o$ and $k_p^e$ are ordinarily polarised and extraordinarily polarised pump wave vectors, respectively. The optic axis of the crystal is marked with 'c' (a). The generated SPDC modes in three dimension showing SPDC annular mode (b).} 
\end{figure}  The condition $\Delta{k_z},\Delta{k_x},\Delta{k_y = 0}$ is known as the condition of perfect phase matching. When this condition is satisfied, the three interacting fields maintain a fixed phase relation leading to most efficient conversion of pump to SPDC photons.For most of the cases, the phase-matching condition is not perfectly satisfied due to pump beam characteristics and nonlinear crystal properties. As a result, a broad spectrum of SPDC photons are generated. To study this, the SPDC photon phase-matching function which is a function of crystal length and pump beam waist is often written as \cite{ramirez2013effects, torres2005spatial}
\begin{align}
  \Phi(\omega_s, \omega_i, \mathbf{k}_s, \mathbf{k_i} ) = E(\mathbf{k}_s+\mathbf{k}_i) sinc\left(\frac{1}{2}L\Delta k(\omega_s, \omega_i,\mathbf{k}_s, \mathbf{k}_i)\right)exp\left(-i\frac{1}{2}L\Delta k(\omega_s, \omega_i, \mathbf{k}_s, \mathbf{k}_i )\right)
\end{align}
where $\omega_s$ and $\omega_i$ are angular frequencies of signal and idler photons, respectively. And $\mathbf{k}_s$, and $\mathbf{k}_i$ are its transverse wave vectors. $E(\mathbf{k}_s+\mathbf{k}_i)$ is the pump transverse mode amplitude distribution, which is a function $\mathbf{k}_s+\mathbf{k}_i$. Longitudinal ($k_{\mu z}$) and transverse wave vectors are related as $k^2_{\mu} = k_{\mu z}^2+ \mathbf{k}_\mu^2$, with $k_\mu = n(\omega_\mu)\omega_\mu/c $, ($\mu$ = p,s,i), where $n(\omega)$ is the refractive index of the medium as a function of signal/idler frequency and c is the velocity of light in vacuum. $L$ is the crystal length. Incorporating equation (1) and (3), the phase mismatch ($\Delta k$) in the equation (4) can be written as
\begin{align}
  \Delta k(w_s, w_i,\mathbf{k}_s, \mathbf{k}_i )=\Delta k_z-\frac{|\mathbf{k}_s+\mathbf{k}_i|^2}{2k_p}+\Delta k_y \tan \rho 
\end{align}
where, $\rho$ is the walk-off angle and $\Delta k_y \tan \rho $ in the equation suggest that there is a walk-off effect in z-y plane. There is no walk-off in the x-z plane. This effect is the reason for the cylindrical asymmetry observed in the y-z plane. In other words, the SPDC mode defined by equation (4) shows a cylindrical asymmetry because of the walk-off effect from equation (5). Since equation (5) is formed from equation (1) and (3) which is developed by considering the geometry of the pump, signal and idler photons with walk-off effect included, it is meaningful to characterise the SPDC mode profile by using coordinate system by considering pump beam walk-off. Therefore, we have developed a mathematical model by using the geometry of three wave vectors shown in Fig. 1(a) to form a coordinate distribution of generated photons. The developed model is,  
\begin{align}
   x = \sin \phi_{\mu} (L-a)\tan\theta_{\mu} 
\end{align}
\begin{align}
   y = \cos \phi_{\mu} (L-a)\tan\theta_{\mu}-a\tan \rho%
\end{align}
Where $x$ and $y$ are coordinate points in the x-y plane. $\phi_\mu$ is the angle of transverse signal/idler wave-vector in the x-y plane for any defined emission angle $\theta_\mu$. Therefore, $\phi_\mu$ takes values between 0 to $2\pi$. $a$ is a variable which takes values between 0 and $L$. Overall, equation (6) and (7) create coordinate points in the x-y plane, and, since $\phi_\mu$ and 'a' are variables it creates a SPDC mode shape as shown in Fig. 1(b). The contribution of the pump waist in the formation of asymmetry due to walk-off is very small compared to crystal length as it will be seen in the result section, therefore, this part is not included in equation (6) and (7). However, one can easily modify the model by considering many pump rays according to pump beam waist. For noncollinear geometry, down-converted photons distributed circularly around the pump respecting the law of momentum conservation. As a result, a conical solid shape of SPDC target modes is formed. As it can be seen from z-direction, an annular distribution of down-converted photons are observed. Due to the walk-off, photons born above the pump beam have a greater separation compared to below ones as shown in Fig.1(a). This effect causes cylindrical asymmetry in the SPDC annulus. However, this separations are equal on the x–z plane. For efficient generation of SPDC photons, condition in equation (1), (2) and (3) should be satisfied. Since the pump beam has a spectral width with definite high intensity central frequency, the condition of perfect phase-matching and phase-mismatching are both seen in a single SPDC experiments. The central frequency of the pump is responsible for the central highest intensity peak in the signal and idler modes. The broad spectrum of the pump, with the phase mismatch responsible for the spectral width in the SPDC modes. \section*{Experiment and Results}
\begin{figure}
\centering\includegraphics[width=0.8\textwidth]{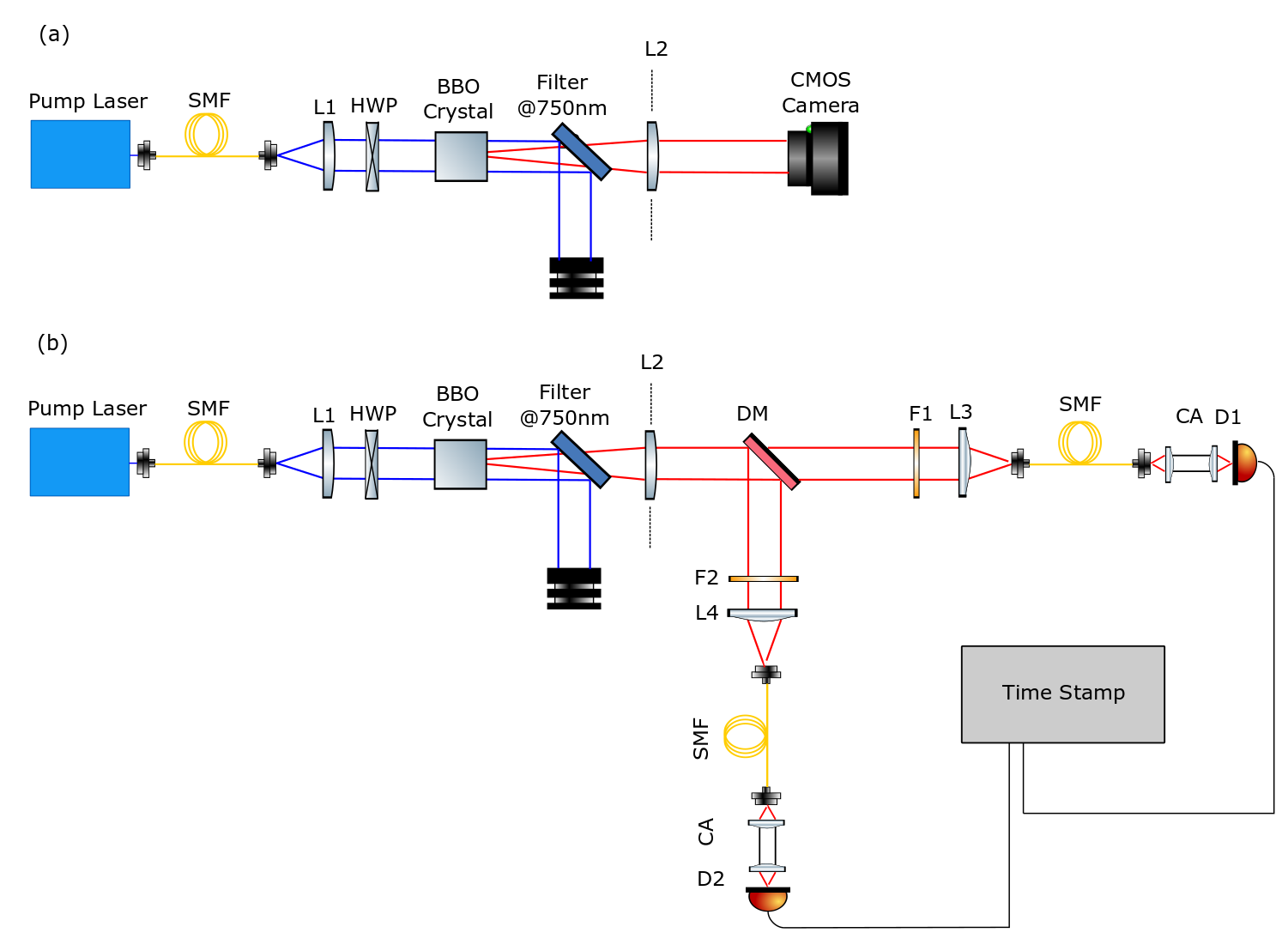}
\caption{Schematic of the experimental configuration for (a) SPDC photon imaging using CMOS camera (b) and coincidence event detection apparatus for signal and idler photons.}
\end{figure}
Our experimental set up is shown schematically in Fig. 2. A collimated Gaussian laser beam, of 40 mW power from a laser diode (LD) centered at 405 nm with bandwidth of 160 MHz, is used as pump. The pump beam is sent through fluorescent filter (not shown) and a half wave plate after mode cleaned using SMF and collimated using a plano-convex lens (L1). Then, the pump illuminates the BBO crystal of which the cut angle is $28.82^{\circ}$. This produces down-converted photons of wavelengths 780 nm and 842 nm of signal and idler, respectively. After the crystal, the pump beam is blocked by using a long-pass filter of cut-on wavelength at 750 nm. Image of remaining signal and idler photons are captured by using CMOS camera (Thorlabs DCC1645C-HQ). We have removed the infrared (IR) filter from the camera because the IR filter cuts off above 650 nm. Later, a laterally movable plano-convex lens (L2) is inserted between the long-pass filter and the camera to study the affect of the L2 on the asymmetry in the SPDC beam profile. In our experiment, five different crystal length values are used; 1 mm, 2 mm, 3 mm, 4 mm, 6 mm and 10 mm. With various combinations of SPDC parameters, we captured the image of SPDC photons. For noncollinear geometry with crystal lengths 10 mm, 6 mm and 4 mm and a corresponding numerical simulation before L2 is inserted is demonstrated in Fig. 3. The observed SPDC modes are seen with asymmetric annular shape which supports the theoretical prediction. We have used equation (6) and (7) to simulate the SPDC images, where it can be seen that our experimental and numerical results are perfectly matching. However, compared to the simulated images, the experimentally captured images have small difference in the region marked with "a" and "b" in Fig. 3. Because, the phase mismatch $(\Delta k)$ arises from finite pump beam waist. However, it is not included in the simulation for the sake of simplicity and it is least interested in the scope of this paper.\begin{figure}[h!]
\includegraphics[width=1\textwidth]{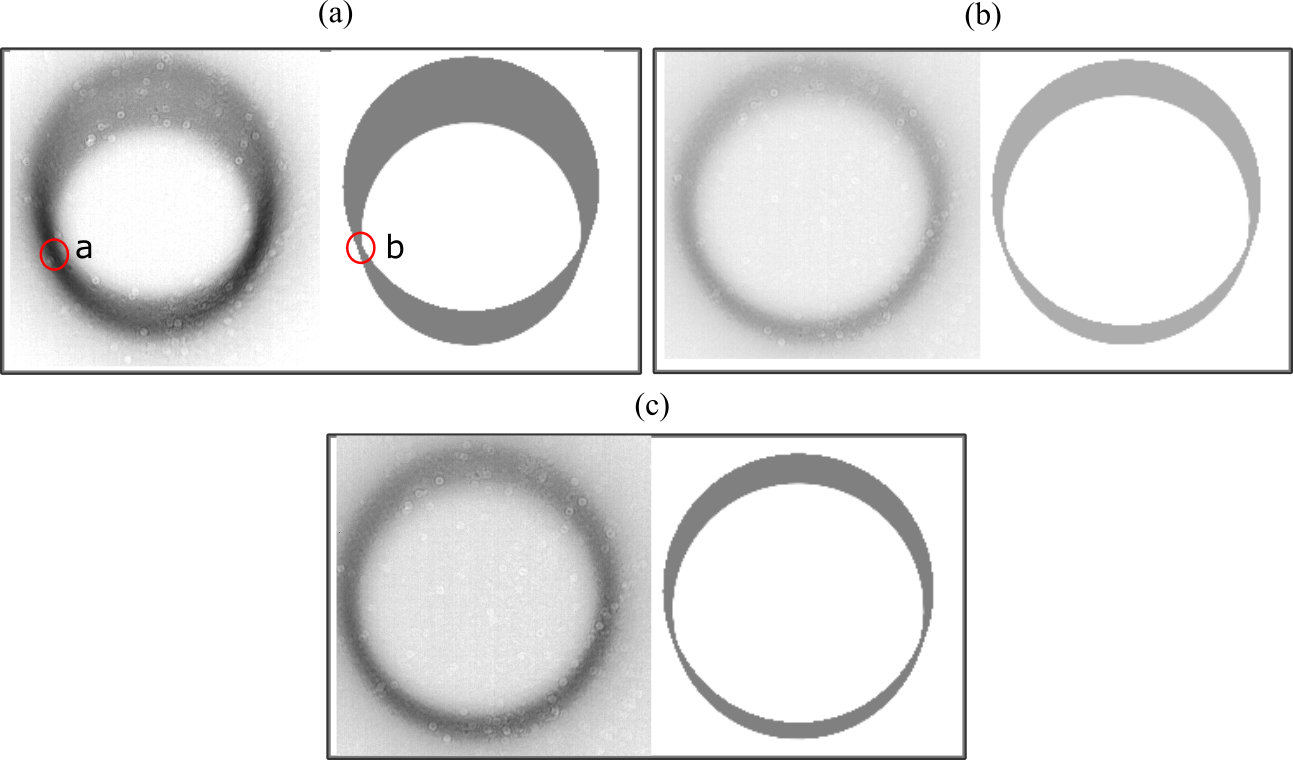}
\centering
\caption{Experimental (left) and the corresponding simulated (right) images of SPDC modes. Panel (a), (b) and (c) are typical asymmetrical profile of the SPDC emission for crystal length values of 10 mm, 6 mm and 4 mm, respectively. Given SPDC modes have $1.5^{\circ}$ emission angle.} 
\end{figure}
Asymmetric factor ($AF$) is introduced to quantify the asymmetry in the SPDC emission profile. It is defined as $AF=1-a/b$, where $a$ is the $1/e^2$ value of the thickness in x axis, and $b$ is that of the y axis. We study the effect of crystal length, pump beam waist and the phase matching angle on AF separately by varying one input parameter and fixing the other two. Firstly, by varying the crystal length we find that the asymmetry of SPDC emission profile initially increases linearly with the length of the crystal and goes flat for larger crystal length as shown in Fig. 4(a). Theoretical plot (solid line) in the figure is plotted by using equation (6) and (7). The error plot in the figure is made by considering the phase mismatch value which is not considered in the equation, along with propagated instrumental errors. Because of the fact that the SPDC photons are generated throughout the length of the crystal, the width of the SPDC annulus is larger for long crystals. This effect, combined with walk-off effect, explained in theory section leads to greater asymmetry in the long crystals compared to short ones. To know the asymmetry due to the alignment of the crystal, we fix the crystal length and pump beam waist, and rotate the BBO crystal effectively changing the emission angle and study its effect on AF. We observed that the asymmetry in the SPDC mode is linearly increasing with emission angle as seen in Fig. 4(b). Because, the walk-off angle for the extraordinarily polarised pump beam with respect to the crystal optic axis increases with emission angle. As a result, asymmetry due to walk-off effect is large for larger emission angles. Three plots in Fig. 4(b) is made with different pump beam waist varied by using biconvex lens before the BBO crystal and carefully focused at the center of the crystal. All three SPDC parameters have important role in deciding down-converted beam profile. Also, it should be noticed that, SPDC photon collection efficiency to SMF is high for large pump beam waist as shown in the literature\cite{zhong201812}. This does not contradict with our result as the affect of pump beam waist is insignificant compared to other effects. Careful design of experimental set is required to neglect the asymmetry in the SPDC profile. However, the complete elimination of the asymmetry is difficult to achieve. 
 
\begin{figure}[h!]
\includegraphics[width=1\textwidth]{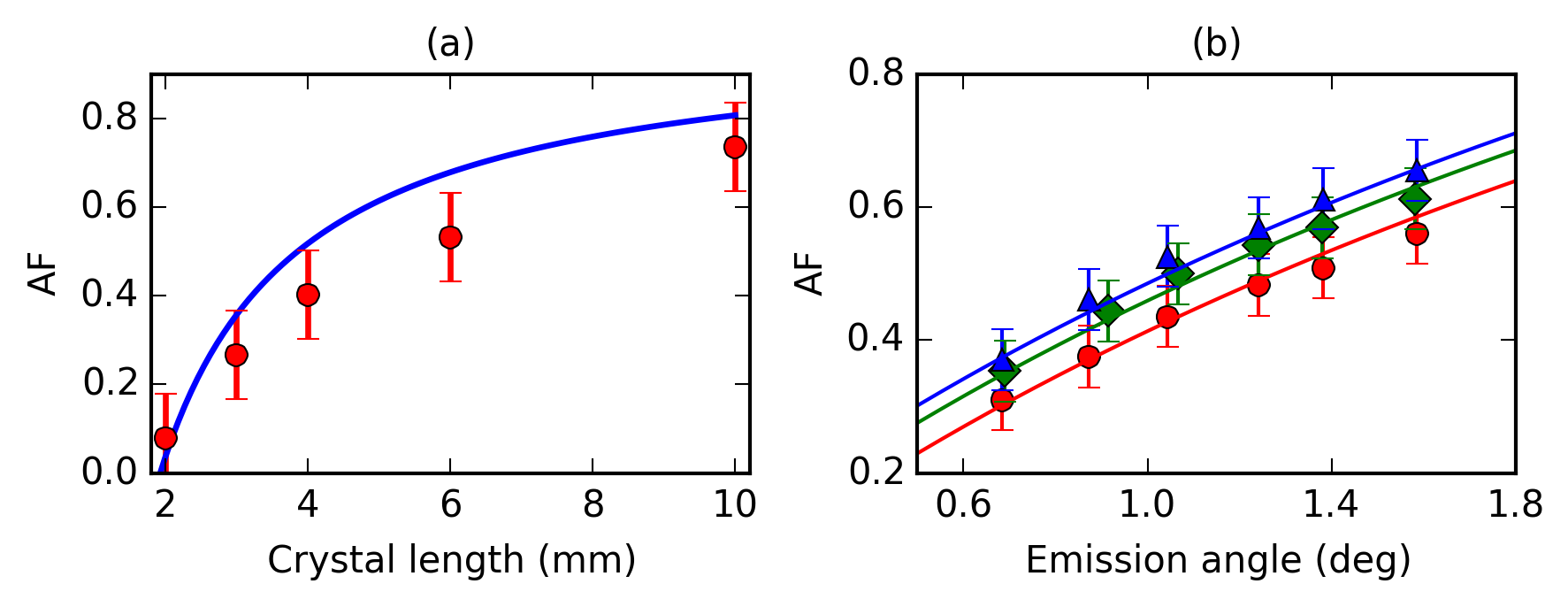}
\centering
\caption{(a)Asymmetry value for crystal length as a varying SPDC parameter. The error value is $\pm 0.1$ in the y-axis. (b) Asymmetry of SPDC photons for different pump beam waist. Triangle, diamond and circle correspond to the pump beam waist of 118 $\mu m$, 78.6 $\mu m$ and 47.2 $\mu m$, respectively. Solid line is the theoretical fit using equation (6) and (7). The propagation of error is calculated to be 0.046 in the y-axis.} 
\end{figure}
we provide a solution to asymmetrical SPDC profile by inserting a lens displaced from optical axis in the lateral direction. Monitoring the SPDC mode profile by tweaking the lateral position of the lens allows the real-time observation of the asymmetry factor. Our method to correct the asymmetry works very effective in improving the SPDC photon coincidence often written as normalized with pump power and time such as brightness (/mW/s). We modified the experimental set up shown in Fig. 2 by introducing a plano-convex, L2, after the long-pass filter. First we observed the noncollinear SPDC mode shape in the camera in the same way previously explained. \begin{figure}[h!]
\includegraphics[width=0.7\textwidth]{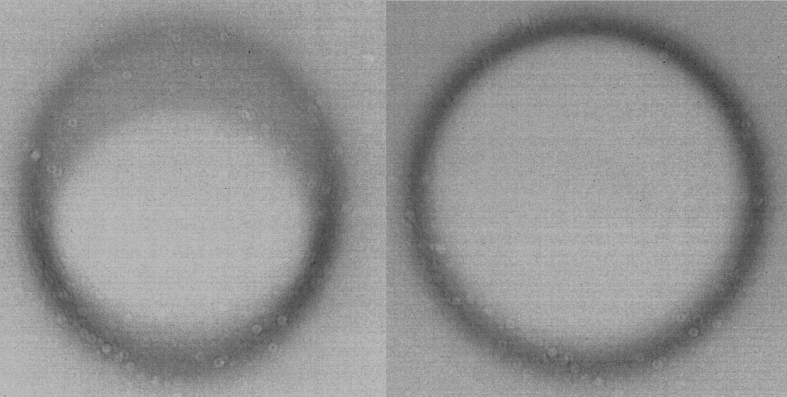}
\centering
\caption{Direct captured image of signal mode (left) and corrected emission asymmetry (right) using plano-convex lens with off-centered positioning.} 
\end{figure}Then, we insert a plano-convex lens with large focal length after the BBO crystal by aligning it at its one focal length apart from the crystal such that conically outward SPDC light is collimated. Afterwards, we moved the lens in x and y axes by observing the live SPDC mode in the camera until the maximum  cylindrical symmetry is achieved. The observed image before and after the correction lens is shown in Fig. 5. It can be seen that the asymmetry is significantly reduced after introducing the lens at the correct position. The picture of the SPDC in collinear geometry is not included as the asymmetry is very difficult to notice. To study the improvement in the brightness of the SPDC photons after introducing the lens, we aligned the crystal for collinear geometry by rotating the BBO crystal and observed the annular mode shape in the camera until it collides into the Gaussian profile.Then, the experimental set up is modified as shown in Fig. 2(b). The down-converted signal and idler beam separated by using a dichroic mirror which differentiate the photon pairs with reference to its wavelength. For mode cleaning, the signal and idler photons are coupled into SMF. Both signal and idler photons travelled in different direction collected separately by using collection arrangement (CA). The CA consists of two aspheric lenses; first one to collimate the diverging photon from SMF and second one to focus this to two detectors, D1 and D1. We have used avalanche photo detectors (APD) which counts the number of photons arrived and registers its arrival time. The D1 and D2 signals are later fed into a time-stamp unit which compares the timing information and process for the coincidence events. 
\begin{figure}[h!]
\includegraphics[width=0.7\textwidth]{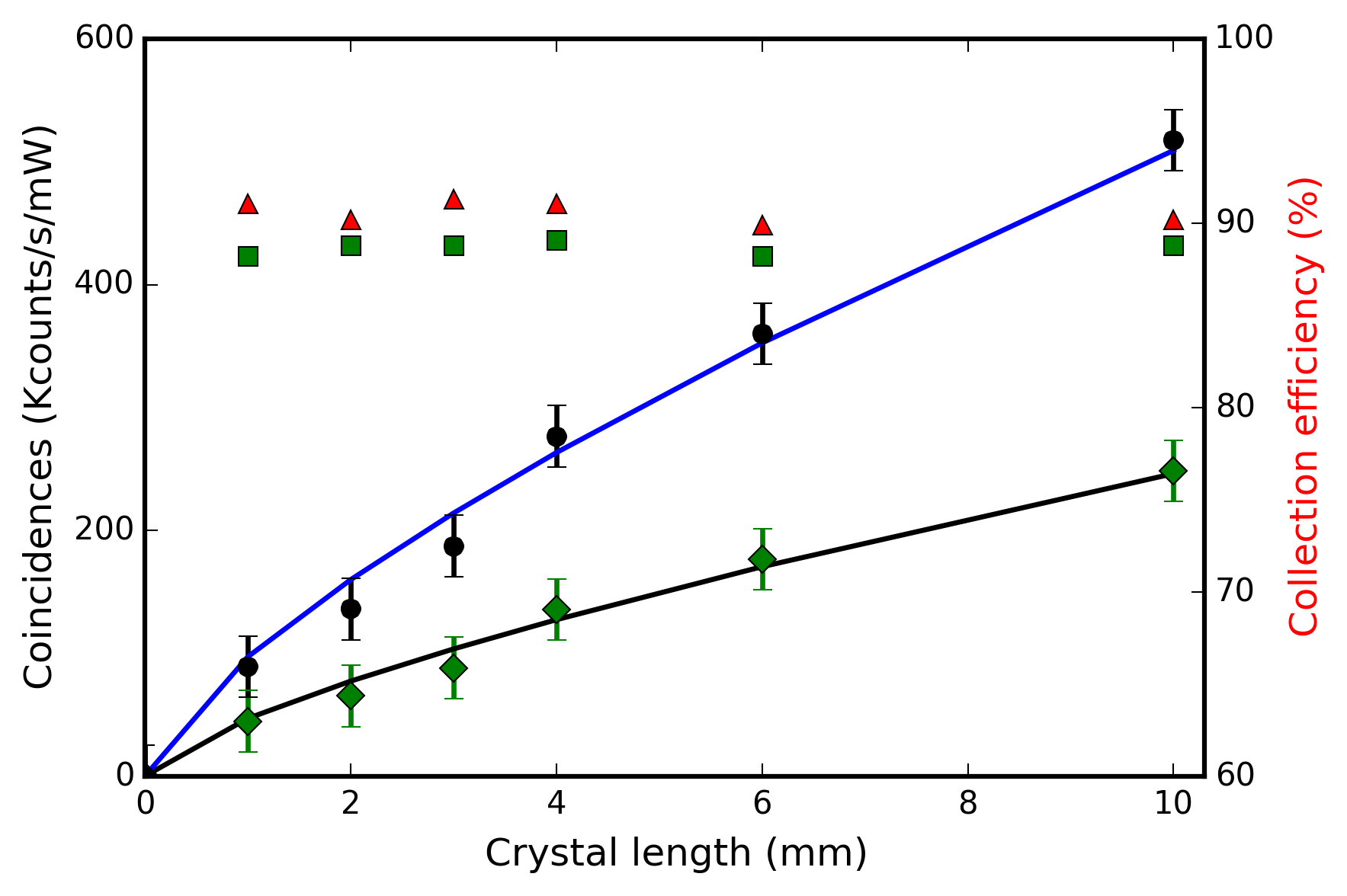}
\centering
\caption{Comparison of SPDC photon coincidence rates for corrected (circle) and uncorrected (diamond) spherical asymmetries which is coupled into SMF. Collection efficiency (right-hand y-axis) for corrected (triangle) and uncorrected (square) asymmetries are shown.} 
\end{figure}

Our data for SPDC photon coincidence rate with and without asymmetry versus different crystal length is shown in Fig. 6. The result shows that there is an improvement of SPDC brightness by a factor of two after correcting the asymmetry observed in the correlated photon experiment. The theoretical fit for the plot is proportional to $L^{0.72}$ which is matching with previous study \cite{trojek2007efficient}. Coefficient of theoretical Fit for corrected case is double of uncorrected ones. A coincidence rate of above 0.5 MCoincidences/mW/s is observed with a 10 mm BBO crystal, which is the highest value achieved in the literature (up to authors' knowledge) for critically phase-matched SPDC sources. A record brightness of 400 KCoincidences/mW/s was previous demonstrated\cite{lohrmann2018high} by using specialized set up with detector efficiency comparable to the one we used($\approx 55\%$). Our achieved brightness is comparable to those in quasi phase-matched sources\cite{steinlechner2012high}.   

Collection efficiency along with the brightness is an important parameter to be considered while designing SPDC photon source. There is a trade-off between these two parameters reported theoretically\cite{} and experimentally as pump beam waist a deciding parameter\cite{}. Larger momentum uncertainty leads to a low collection efficiency. While photon flux density defines the brightness of the SPDC source. Therefore, larger pump beam waist is desirable for high collection efficiency by sacrificing SPDC brightness as large pump beam waist leads to low photon flux density. Spacial distribution of pump beam such as angle of the pump and its waist size inside the crystal decides the collection efficiency. Therefore, with fixed spacial distribution and increased crystal length does not change the momentum uncertainty hence collection efficiency. However, the photon brightness is increased with crystal length as photon flux density is increased with crystal length. In another word, the crystal length creates more area of cross-section to generate more photon pairs resulting high brightness, since number of singles also increased along with crystal length collection efficiency is not changed. Our results support this claim consistent with previous study\cite{oi2017cubesat}. The collection efficiency for signal/idler photons is defined by $\mu_{s(i)} = \frac{C}{s_{i(s)}d_st_s}$ 
where $C$ is the number of coincidence events, $s_i(s)$ is the number of registered singles event in the idler/signal channel. $d_{s(i)}$ is the signal/idler detector efficiency. $t_{s(i)}$ is the transmission probability in signal/idler channel. the combined collection efficiency can be found by taking geometric mean of the collection efficiency of signal and idler photons, that is $\sqrt{\mu_{s}\mu_{s}}$. Our detector quantum efficiency is $55\%$ at signal channel and $50\%$ for idler. To find the $t_{s(i)}$ in each channels, we multiplied transmission coefficient of each optical components in the respective channels. Long pass filter optical fibers are the most lossy components with $0.88\%$ transmission coefficient each. The total $t_{s(i)}\approx0.645$ for each channel. With $L=10 mm$, we found collection efficiency to be $90.20\%$ and $88.79\%$ for corrected and uncorrected cases, respectively. These values shows an improvement from previous studies for this type of system\cite{septriani2016thick}.   
\section*{Conclusion}
In this paper we used CMOS camera, which is first time, to study the SPDC mode profile. We also observed SPDC profile modes with Sony PS3 camera, which is a CCD camera. We preferred using Thorlab's CMOS camera in the experiment because of its better user interface. The use of such inexpensive and easily available cameras is a significant advantage over intensified charge coupled device (ICCD or EMCCD) cameras, which are few order of magnitude more expensive. The fact that single photon detection sensitivity is not needed allow us to use any camera that have sufficient responsivity in the near-infrared range to capture SPDC modes with reasonable integration times. 

In noncollinear geometry, our captured image shows an asymmetry in the mode shape. This is because of walk-off effect in the nonlinear crystal. Our result concluded that input parameters such as pump beam waist and crystal length affect phasematching condition leading to changing the down-converted photon mode characteristics. Our result shows, larger pump beam waist and crystal length have high asymmetry. Additionally, phasematching angle which arise from the nonlinear crystal orientation defines the degree of walk-off hence affect the AF; which we introduced to characterise the asymmetry. We have provided a mathematical model to characterise the obtained SPDC mode. Our mathematical model is simple to program and sufficient to recreate the observed mode. This will open further research in SPDC mode characterisation.

Characterising SPDC sources is paramount importance since it is used in several quantum mechanics applications. Brightness of the source and collection efficiency are two important parameters to qualify the SPDC source. We have demonstrated a recorded SPDC source brightness with improved collection efficiency using critically phase matched type-I nonlinear crystal. This could be achieved by correcting the asymmetry in the SPDC mode profile by using plano-convex lens. The demonstrated theoretical model and experiment are very useful in efficient demonstration of SPDC source.   

\section*{Acknowledgments}
Authors thank A. Dandasi and C. Dindar for their contribution on the preparation of the paper. This work was supported  by Scientific and Technological Research Council of Turkey (TUBITAK) with project number 118E156.
\bibliography{references.bib}

\end{document}